\documentclass[aps,prl,twocolumn,superscriptaddress,amsmath,amssymb]{revtex4-2}
\usepackage{cancel}

\usepackage{graphicx}
\usepackage{dcolumn}
\usepackage{bm}
\usepackage{hyperref}
\usepackage[vcentermath, enableskew]{youngtab} 
\usepackage{hyphenat}
\usepackage{amsthm}
\theoremstyle{plain}

\theoremstyle{remark}

\theoremstyle{definition}

\newcommand{\SU}{\ensuremath{\mathrm{SU}}}

\newcommand{\idm}{\ensuremath{1\!\!1}}

\newcommand{\A}{\ensuremath{\mathrm{A}}}
\newcommand{\B}{\ensuremath{\mathrm{B}}}
\newcommand{\na}{\ensuremath{n_{\mathrm{A}}}}
\newcommand{\nb}{\ensuremath{n_{\mathrm{B}}}}
\newcommand{\la}{\ensuremath{\lambda^{(\mathrm{A})}}}
\newcommand{\lb}{\ensuremath{\lambda^{(\mathrm{B})}}}
\newcommand{\lab}{\ensuremath{\lambda^{(\mathrm{AB})}}}

\newcommand{\AB}{\ensuremath{\mathrm{AB}}}

\begin{document}


\title{Extendibility of Werner States}

\author{Dávid Jakab}
\email{jakab.david@wigner.hu}
\affiliation{Wigner Research Centre for Physics,  H-1525 Budapest P.O. Box 49, Hungary}
\affiliation{University of Pécs, Ifjúság utca 6, 7624 Pécs, Hungary}
\author{Adrian Solymos}%
\email{solymos.adrian@wigner.hu}
\affiliation{Wigner Research Centre for Physics,  H-1525 Budapest P.O. Box 49, Hungary}
\affiliation{Eötvös Loránd University, Pázmány Péter sétány 1/A 1117 Budapest, Hungary}

\author{Zoltán Zimborás}%
\email{zimboras.zoltan@wigner.hu}
\affiliation{Wigner Research Centre for Physics,  H-1525 Budapest P.O. Box 49, Hungary}
\affiliation{Eötvös Loránd University, Pázmány Péter sétány 1/A 1117 Budapest, Hungary}
\affiliation{Algorithmiq Ltd, Kanavakatu 3 C, FI-00160 Helsinki, Finland}
%



\begin{abstract}
  We investigate the two-sided symmetric extendibility problem of Werner states. The interplay of the unitary symmetry of these states and the inherent bipartite permutation symmetry of the extendibility scenario allows us to map this problem into the ground state problem of a highly symmetric spin-model Hamiltonian. We solve this ground state problem analytically by utilizing the representation theory of \(\SU(d)\), in particular a result related to the dominance order of Young diagrams in Littlewood-Richarson decompositions.
  As a result, we obtain necessary and sufficient conditions for the extendibility of Werner states for arbitrary extension size and local dimension. Interestingly, the range of extendible states has a non-trivial trade-off between the extension sizes on the two sides. We compare our result with the two-sided extendibility problem of isotropic states, where there is no such trade-off.

\end{abstract}

\maketitle


\textit{Introduction.}---State extension problems have played a prominent role in entanglement and non-locality theory since the very beginning~\cite{werner1989application, werner1990remarks,doherty02_distin_separ_entan_states, terhal2003symmetric,terhal2004entanglement, christandl2004squashed, doherty04_compl_famil_separ_criter}. The so-called {\it symmetric extendibility\/} (or {\it shareability\/}) turned out to be a particularly useful concept, that captures how much a bipartite state can be shared
between parties: A bipartite state between Alice and Bob is said to be $(\na,\nb)$-extendible if there exists a state between $\na$ number of Alices and $\nb$ number of Bobs, such that the reduced state of each pair is the original one. This notion, in both its one-\cite{werner1989application, werner1990remarks,terhal2004entanglement, doherty2014entanglement} and two-sided versions~\cite{doherty02_distin_separ_entan_states, terhal2003symmetric,johnson13_compat_quant_correl}, originally arose in the context of the characterization of the set of entangled and non-local states. In particular, it was shown that only separable states are arbitrarily, or \((\infty, \infty)\)-extendible~\cite{fannes88_symmet_states_compos_system, doherty02_distin_separ_entan_states}, however, bounding quantitatively the distance to separable states through the degree of extenbility was done much later~\cite{doherty2014entanglement,brandao2017quantum,brandao2011faithful} using novel versions of the Quantum de Finetti theorem~\cite{konig2005finetti,christandl2007one}. It was also shown, that certain highly extendible states that are still far away in trace norm from the set of separable states are useful for quantum data-hiding~\cite{brandao2011faithful}.
Additionally, symmetric extendibility turned out to be also a key concept in quantum key distribution~\cite{PhysRevA.74.052301,PhysRevA.79.042329,khatri2016symmetric}. The degree of extendibility itself is an entanglement monotone~\cite{nowakowski2016symmetric}\footnote{Currently only monotonicity for 1-LOCC operations is proven rigorously.}, but it can also be related to entanglement measures such as the family of measures called unextendible entanglement~\cite{wang2019quantifying}, or the squashed entanglement for which it serves as a lower bound~\cite{brandao2011faithful,li2018squashed}. Recently also a complete resource theory was developed based on the notion of symmetric extendibility~\cite{PhysRevLett.123.070502}.

Despite the prominent role extendibility plays in entanglement theory, it has been calculated analytically only for a few families of entangled states~\cite{lami2019extendibility,Ranade_2009}.
When discussing entanglement properties of states, it has been useful to consider examples of entangled states with high symmetry, as the symmetry could be used to greatly simplify the computation of entanglement measures. The most notable such states are Werner states~\cite{PhysRevA.40.4277, keyl2002fundamentals}, for which several entanglement measures have been determined~\cite{vollbrecht01_entan_measur_under_symmet}. Naturally, also the symmetric extendibility problem of Werner states has attracted attention. 
An analytic condition for one-sided, i.e., $(1,n)$-extendibility was previously derived, but as of now, the general two-sided problem has only been solved for a couple of specific extension sizes~\cite{johnson13_compat_quant_correl}. In this paper, we close this gap and obtain necessary and sufficient conditions in an analytic form for the extendibility of Werner states for arbitrary extension size and local dimension.

 \textit{Extendibility.}---We call a state $\rho$ on $\mathcal{H}_\A\otimes\mathcal{H}_\B$ $(\na,\nb)$-extendible, if there exists a state $\hat{\rho}$ on  ${\left(\mathcal{H}_\A\right)}^{\otimes \na}\otimes{\left(\mathcal{H}_\B\right)}^{\otimes \nb}$ such that for all $i=1,2, \ldots ,\na$ and $j=1,2, \ldots,\nb$,
 \begin{equation}
   \mathrm{Tr}_{\overline{\A}_{i}\overline{\B}_{j}}\left (\hat{\rho} \right )=\rho\text{,}
 \end{equation}
 where \(\mathrm{Tr}_{\overline{\A}_{i}\overline{\B}_{j}}\) denotes the partial trace that restricts to the $i$-th of Alice's and $j$-th of Bob's Hilbert spaces. Some authors require permutation symmetry within the Alice's and Bob's\ subsystems; however, let us note, that for any \((\na,\nb)\)-extendible state, one can obtain an extension with such bipartite permutation symmetry by twirling an arbitrary \((\na,\nb)\)-extension with the permutation groups, $S_{\na} \times S_{\nb}$, of Alice's and Bob's subsystems.

 Clearly, each \((\na,\nb)\)-extendible state is \((n'_{\A},n'_{\B})\)-extendible for for all \(n'_{\A}\le \na\) and \(n'_{\B}\le \nb\); this motivates a partial order on \((\na,\nb)\) pairs. As only separable states can be extended infinitely, all entangled states can be described by a set of maximal extenibilities w.r.t.~this partial order. For example, in the case of pure entangled states the single maximal extendibility \((1,1)\).
 


 \textit{Werner and isotropic states.}---In general, calculating the maximal extendibility numbers for an arbitrary state is difficult and can only be done numerically~\cite{doherty02_distin_separ_entan_states,doherty04_compl_famil_separ_criter, PhysRevLett.103.160404,boyd04_convex, benson00_solvin_large_scale_spars_semid}. Restricting the analysis to highly symmetric states allows us to use representation theoretic techniques to derive analytic results.
 The most well-known  types of  states with such symmetries are Werner and isotropic states, invariant to local unitary transformations of the form  $U \otimes U$ and $U\otimes \overline{U}$ respectively. Both are one-parameter families of states; the former is commonly parametrized by the expected value of the flip operator \(F\), while the latter by its partial transpose:
\begin{align}
    & \rho^{\mathrm{W}}(\alpha)=\frac{d}{d^2-1}\left[(d-\alpha) \frac{\idm}{d^2}+\left(\alpha - \frac{1}{d}\right) \frac{F}{d}\right]\text{,}\\
    &\rho^{\mathrm{I}}(\beta)=\frac{d}{d^2-1}\left[(d-\beta) \frac{\idm}{d^2}+\left(\beta - \frac{1}{d}\right) \frac{F^{\mathrm{t_\B}}}{d}\right]\text{,}
\end{align}
%
where $\alpha \in [-1,1]$ and $\beta \in [0,d]$ are the expected values \(\alpha=\mathrm{Tr}(\rho^{\mathrm{W}}(\alpha)F)\) and \(\beta=\mathrm{Tr}(\rho^{\mathrm{I}}(\beta)F^{t_{\B}})\). Werner states are separable iff $0\leq \alpha$, while isotropic states iff $\beta \leq 1$. As the $(\na,\nb)$-extendible Werner states form a convex set, they correspond to a parameter interval \([\alpha_{\na,\nb},1]\). Therefore, the extendibility problem reduces to finding, for each fixed \((\na,\nb)\) pair, the parameter \(-1\le\alpha_{\na,\nb}<0\) that corresponds to the most entangled, \((\na,\nb)\)-extendible Werner state.





\textit{Extendibility as a ground state problem.}---Following Ref.~\cite{johnson13_compat_quant_correl}, we show that finding the most entangled, \((\na,\nb)\)-extendible Werner state is equivalent to solving the ground state problem of a certain spin Hamiltonian.
All composite states that are invariant to local unitary transformations and the bipartite permutations of \(S_{\na}\times S_{\nb}\), are \((\na,\nb)\)-extensions of some Werner states. Conversely, the twirl of a Werner state's \((\na,\nb)\)-extension with the two previous groups, is an \((\na,\nb)\)-extension of the same Werner state; thus, all extendible Werner states have unitary and bipartite permutation symmetric extensions. Consider the ``Hamiltonian'',
\begin{equation}
  \label{eq:1}
  H^{\mathrm{W}}=\frac{1}{\na\nb}\sum_{i\in L,j\in R}F_{ij}.
\end{equation}
The normalized eigenprojectors of \(H^{W}\) are \((\na,\nb)\)-extensions of certain Werner states; moreover, since the flip operator plays an important role in our parametrization, if \(\hat{\rho}^{\mathrm{W}}(\alpha)\) is an \((\na,\nb)\)-extension of the Werner state \(\rho^{\mathrm{W}}(\alpha)\), then \(\text{Tr}(\hat{\rho}^{\mathrm{W}}(\alpha)H^{\mathrm{W}})=\alpha\). Consequently, the smallest eigenvalue of \(H^{\mathrm{W}}\) must be equal to \(\alpha_{\na,\nb}\), as the existence of an \((\na,\nb)\)-extension for a Werner state with parameter \(\alpha<\min\text{Spect}(H^{\mathrm{W}})\) would lead to contradiction.

The interplay of the permutation and unitary symmetries of \(H^{\mathrm{W}}\) allows us to re-express it in terms of tensor product representations of a Casimir operator of \(\SU(d)\). For a detailed explanation, see Section A of the supplemental material. 
\begin{equation}
  \label{eq:3}
  H^{\mathrm{W}}=\frac{1}{2\na\nb}\left( C_{\AB}-C_{\A}-C_{\B}\right)+\frac{\idm}{d},
\end{equation}
where \(C_{\A}\), \(C_{\B}\) and \(C_{\AB}\) denote the quadratic Casimir operator of \(\SU(d)\) in the \(\na\), \(\nb\) and \(\na+\nb\)-fold tensor products of the defining representation, acting on Alice's and Bob's subsystems and the entire system respectively.

By an analogous argument, one may also devise a spin Hamiltonian which has a ground state energy equal to the parameter of the most entangled isotropic state, \(1<\beta_{\na,\nb}\le d\). The main difference compared to Eq.~\eqref{eq:3} is that in this case, the dual, i.e.~the complex conjugate of the defining representation appears on Bob's subsystems:
\begin{equation}\label{eq:4}  H^{\mathrm{I}}=\frac{1}{2\na\nb}\left(\tilde{C}_{\AB}-C_{\A}-C_{\B}\right)-\frac{\idm}{d},
\end{equation}
where  \(\tilde{C}_{\AB}\) denotes the quadratic Casimir operator in the \(\na+\nb\)-fold tensor product representation in which \(U\in\SU(d)\) is represented as itself on Alice's subsystem, and as \(\overline{U}\) on Bob's subsystem.

Finally, we note that \(H^{\mathrm{W}}\) and \(H^{\mathrm{I}}\) may in fact be interpreted as many-body Hamiltonians. The tensor product representations of Casimir operators that appear in them describe permutation symmetric, \(d\)-level magnetic systems that interact with \(\SU(d)\) symmetric exchange interaction. This makes \(H^{\mathrm{W}}\) the Hamiltonian of a \(d\)-dimensional generalization of the bipartite permutation symmetric spin system investigated in~\cite{jakab_2021_quant_phases}. Using the ground state energy derived in the paper, we immediately obtain that for \(d=3\), \(\alpha_{n,n}=-1/n\).

\textit{Solving the ground state problem.}---Since the eigenspaces of the Casimir operators are the \(\SU(d)\) irreducible subspaces, we are able to use representation theory to deal with the eigenproblems of \(H^{\mathrm{W}}\) and \(H^{\mathrm{I}}\). We label the irreducible representations (irreps) of \(\SU(d)\) with Young diagrams of at most \(d\) rows, i.e.~integer partitions with at most \(d\) elements. This way, the eigenvalues of a Casimir operator in an \(n\)-fold tensor product representation correspond to all \(d\)-row Young diagrams \(\lambda\) with \(n\) boxes, or all \(d\)-partitions of \(n\); we denote this as \(\lambda\vdash_{d}n\). In the following, we will conflate integer partitions, Young diagrams, and even \(\SU(d)\) irreps in our notation whenever it does not cause a misunderstanding.

There is an additional intricacy of this labeling scheme that we must take note of: The mapping between the irreps of \(\SU(d)\) and Young diagrams is not a bijection, every irrep of \(\SU(d)\) corresponds to an equivalence class of Young diagrams. Two diagrams belong to the same equivalence class iff they differ in columns of height \(d\), i.e, \(\lambda\cong\lambda'\) iff for some \(M\in\mathbb{Z}\), \(\lambda_{i}=\lambda'_{i}+M\) for all \(i=1,2, \ldots,d\). See, e.g.,~\cite{juergen97_symmet} for an explanation. This means, that two different partitions of the same number cannot belong to the same equivalence class.

Now we are able to introduce the relationship between the labels of an irrep and its dual. A diagram in the equivalence class that correspond to the irrep \(\overline{\lambda}\) has row lengths \(\overline{\lambda}_{i}=M-\lambda_{d-i+1}\), for some fixed \(M\in\mathbb{N}\) such that \(M\ge \lambda_{1}\). One can visualize this as the diagram that complements \(\lambda\) to a height \(d\), width \(M\) rectangle (the \(\SU(d)\) singlet), rotated by \(\pi\). 

Since the Casimirs that appear in \(H^{\mathrm{W}}\) and \(H^{\mathrm{I}}\) commute with each other, the eigenvalues of these Hamiltonians are labeled by triples of Young diagrams, \((\la,\lb,\lab)\), that correspond to the irreps appearing on Alice's and Bob's subsystems and the entire system respectively. Not all possible triples correspond to existing eigenvalues however. The condition for compatibility is, in the case of \(H^{\mathrm{W}}\), that the irrep \(\lab\) must appear in the irrep decomposition of  the tensor product \(\la\otimes \lb\); and in the case of \(H^{\mathrm{I}}\), that \(\lab\) must appear in the decomposition of \(\la\otimes \overline{\lambda}^{(\B)}\), where \(\overline{\lambda}^{(\B)}\) denotes the dual representation of \(\lambda^{(R)}\).

We deal with the constraints on \(\lab\) in the ground state problem, by solving another problem related to the product of irreps: Out of the \(\SU(d)\) irreps that appear in the irrep decomposition of \(\la\otimes \lb\), which one(s) corresponds to the lowest eigenvalue of the quadratic Casimir operator? The knowledge of the solution, as a function of \(\la\) and \(\lb\), reduces the variables of the ground state problem to the two partitions \(\la\) and \(\lb\), for which the only remaining constraint is that they must be d-partitions of \(\na\) and \(\nb\).

An integer partition \(\lambda\vdash n\) is said to dominate \(\mu\vdash n\), which we denote as \(\lambda\unrhd\mu\), if
\begin{equation}
  \lambda_{1}+\lambda_{2}+ \cdots+\lambda_{i}\ge\mu_{1}+\mu_{2}+ \cdots+\mu_{i},
\end{equation}
for all \(i\ge 1\); or equivalently, if one can obtain the diagram of \(\lambda\) from that of \(\mu\) by only moving boxes upwards (with no regard for their horizontal position). This defines a partial order between the integer partitions of \(n\), e.g., \((3,3)\unrhd (2,2,1,1)\), but \((3,3)\) and \((4,1,1)\) are not related. The diagram \(\lambda\) covers \(\mu\) in dominance order, i.e.~\(\lambda\rhd\mu\) and there is no intermediary diagram \(\nu\) such that \(\lambda\rhd\nu\rhd\mu\), iff one can obtain the diagram \(\lambda\) from that of \(\mu\), by removing a single box from the end of some row \(k\), and appending it to row \(i<k\), where after the box is removed from \(\mu\), the rows \(i\) through \(k\) of both diagrams all have the same length, but row \(i-1\) is different. E.g.,
\begin{equation}
  \yng(4,3,2)\quad\text{covers}\quad\yng(3,3,3)\ .
\end{equation}
For a proof see Ref.~\cite{brylawski1973lattice}.
The solution to our intermediary problem starts with a key observation about the eigenvalues of the quadratic Casimir operators,
\begin{equation}
  \label{eq:5}
  c(\lambda)= \sum_{i=1}^d \left[ {\left( \lambda_{i}-\frac{n}{d}\right)}^{2}+2(d-i)\left( \lambda_{i}-\frac{n}{d} \right) \right],
\end{equation}
  where \(n = \sum_{i=1}^d \lambda_{i}\).
  Let \(\lambda\vdash n\) cover \(\mu\vdash n\), and let the indices \(k\) and \(i\) be such that in the previous paragraph, then,
\begin{multline}
  \label{eq:6}
  c(\lambda)-c(\mu)=(k-i)+(\lambda_{i}-\lambda_{k})-2=\\(k-i)+(\mu_{i}-\mu_{k})\ge 1.
\end{multline}
This means, that the order of the eigenvalues of the quadratic Casimir operators, is a refinement of the dominance order. Thus, if the set of irreps that appear in the irrep decomposition of \(\la\otimes \lb\) has a minimum w.r.t.~dominance order, then this minimum must correspond to the eigenvalue of the quadratic Casimir that is smallest among the product diagrams.

The decomposition of a product of \(\SU(d)\) irreps, \(\la\otimes\lb\) is governed by the Littlewood-Richardson rule~\cite{richardson_1934_group_charac_algeb}. This is a combinatorial algorithm, that involves listing all the ways one can attach  the boxes of \(\lb\) as a Young diagram to \(\la\), subject to certain restrictions. Using a result from~\cite{azenhas99_admis_inter_invar_factor_produc_matric}, about the connection of the Littlewood-Richardson algorithm to dominance order, in Section B of the supplemental material, we show that the irrep decomposition of \(\la\otimes \lb\) always has a minimum, \(\hat{\lambda}^{(\AB)}(\la,\lb)\),  w.r.t.~dominance order,
\begin{equation}
  \label{eq:7}
  \hat{\lambda}^{(\AB)}(\la,\lb)=\text{sort}{\{\la_i+\lb_{d-i+1}\}}_{i=1}^{d}.
\end{equation}
That is, to obtain the minimum diagram, one has to attach the rows of \(\lb\), turned upside down, to \(\la\), and sort the rows of the resulting diagram  in decreasing order by their length.

\begin{figure}[ht]
  \includegraphics[width=0.9\columnwidth]{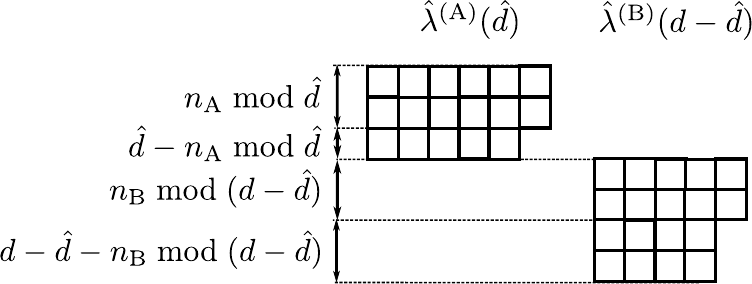}
  \caption{\label{fig:wernersolmaintext}An example for the pair of diagrams that minimizes \(E^{W}(\la,\lb)\). The source of the modulos appearing in Eq.~\eqref{eq:11} is, that it may not be possible to partition \(\na\) into \(\hat{d}\) and \(\nb\) into \(d-\hat{d}\) equal integers, in which case \(\hat{\lambda}^{(\A)}(\hat{d})\) and \(\hat{\lambda}^{(\B)}(d-\hat{d})\) take the most ``rectangle-like'' form available.}
\end{figure}

\textit{The extendibility of Werner states.}---The knowledge of the minimum product diagram, Eq.~\eqref{eq:7}, simplifies the ground state problem of \(H^{\mathrm{W}}\) enough that we are able to solve it analytically. The eigenvalue of \(H^{\mathrm{W}}\) corresponding to the triple \((\la,\lb,\hat{\lambda}^{(\AB)}(\la,\lb))\) is,
\begin{equation}
  \label{eq:8}
  \begin{split}
    &E^{\mathrm{W}}(\la,\lb)=\\
    &\frac{1}{2\na\nb}\left[c(\hat{\lambda}^{(\AB)}(\la,\lb))-c(\la)-c(\lb)\right]+\frac{1}{d}=\\
    &\frac{1}{\na\nb}\sum_{i=1}^d \left[ \la_{i}\lb_{d-i+1}-i(\text{sort}{({\{\la_{j}+\lb_{d-j+1}\}}_{j=1}^{d})}_{i}\right.-\\
    &\hspace{6cm}\left.(\la_{i}+\lb_{i})) \right].
  \end{split}
\end{equation}
In order obtain \(\alpha_{\na\nb}\), we have to minimize \(E^{\mathrm{W}}\) subject to the constraints \(\la\vdash_{d}\na\) and \(\lb\vdash_{d}\nb\). It is possible to choose \(\la\) and \(\lb\) in such a way, that \(\la_{i}\) and \(\lb_{d-i+1}\) cannot be simultaneously non-zero, i.e., the value of the quadratic term in Eq.~\eqref{eq:8} is zero. It is reasonable to expect that, at least for sufficiently large \(\na\) and \(\nb\), the minimum of \(E^{\mathrm{W}}\) corresponds to such a non-overlapping pair of diagrams. Guided by this intuition, in Section C of the supplemental material we show that this is indeed the case not just for large, but for arbitrary values of \(\na\) and \(\nb\). In fact, among the pairs of diagrams that minimize \(E^{W}\), there has to be at least one consisting of certain special, non-overlapping shapes:
\begin{equation}
  \label{eq:9}
  \min_{\substack{\la\vdash_{d}\na,\\\lb\vdash_{d}\nb}}E^{\mathrm{W}}(\la,\lb)=E^{\mathrm{W}}(\hat{\lambda}^{(\A)}(\hat{d}),\hat{\lambda}^{(\B)}(d-\hat{d})),
\end{equation}
for at least one value of \(\hat{d}=1,2, \ldots d-1\) where,
\begin{equation}
  \label{eq:10}
  \begin{split}
    &\hat{\lambda}^{(\A)}(d_\A)=
    \left(\left\lceil \frac{\na}{d_{\A}}\right\rceil, \left\lceil \frac{\na}{d_{\A}} \right\rceil, \ldots , \overset{\na\bmod d_\A\text{-th}}{\left\lceil \frac{\na}{d_\A}\right\rceil} \right.,\\
    &\qquad\qquad\quad \left.\left\lfloor \frac{\na}{d_\A} \right\rfloor,\left\lfloor \frac{\na}{d_\A} \right\rfloor, \ldots , \overset{d_\A\text{-th}}{\left\lfloor \frac{\na}{d_\A} \right\rfloor}\right),\\
    &\hat{\lambda}^{(\B)}(d_{\B})=\left(\left\lceil\frac{\nb}{d_\B}\right\rceil,\left\lceil\frac{\nb}{d_\B}\right\rceil, \ldots , \overset{\nb\bmod d_\B\text{-th}}{\left\lceil\frac{\nb}{d_\B}\right\rceil}\right.,\\
    &\qquad\qquad\quad\left.\left\lfloor \frac{\nb}{d_\B} \right\rfloor,\left\lfloor \frac{\nb}{d_\B} \right\rfloor, \ldots,\overset{d_\B\text{-th}}{\left\lfloor \frac{\nb}{d_\B} \right\rfloor}\right),
  \end{split}
\end{equation}
and \(\lceil . \rceil\) and \(\lfloor . \rfloor\) denote the ceiling and floor functions respectively.
In other words, at least one pair of diagram that minimizes \(E^{\mathrm{W}}\) corresponds to a bipartition of the dimension \(d\) into \(\hat{d}\) and \(d-\hat{d}\), and the elements of the pair are the minima of the sets of all \(\hat{d}\)-partitions of \(\na\) and all \(d-\hat{d}\) partitions of \(\nb\)  w.r.t.~dominance order,  see Fig.~\ref{fig:wernersolmaintext}.

\begin{figure}[ht]
  \includegraphics[width=\columnwidth]{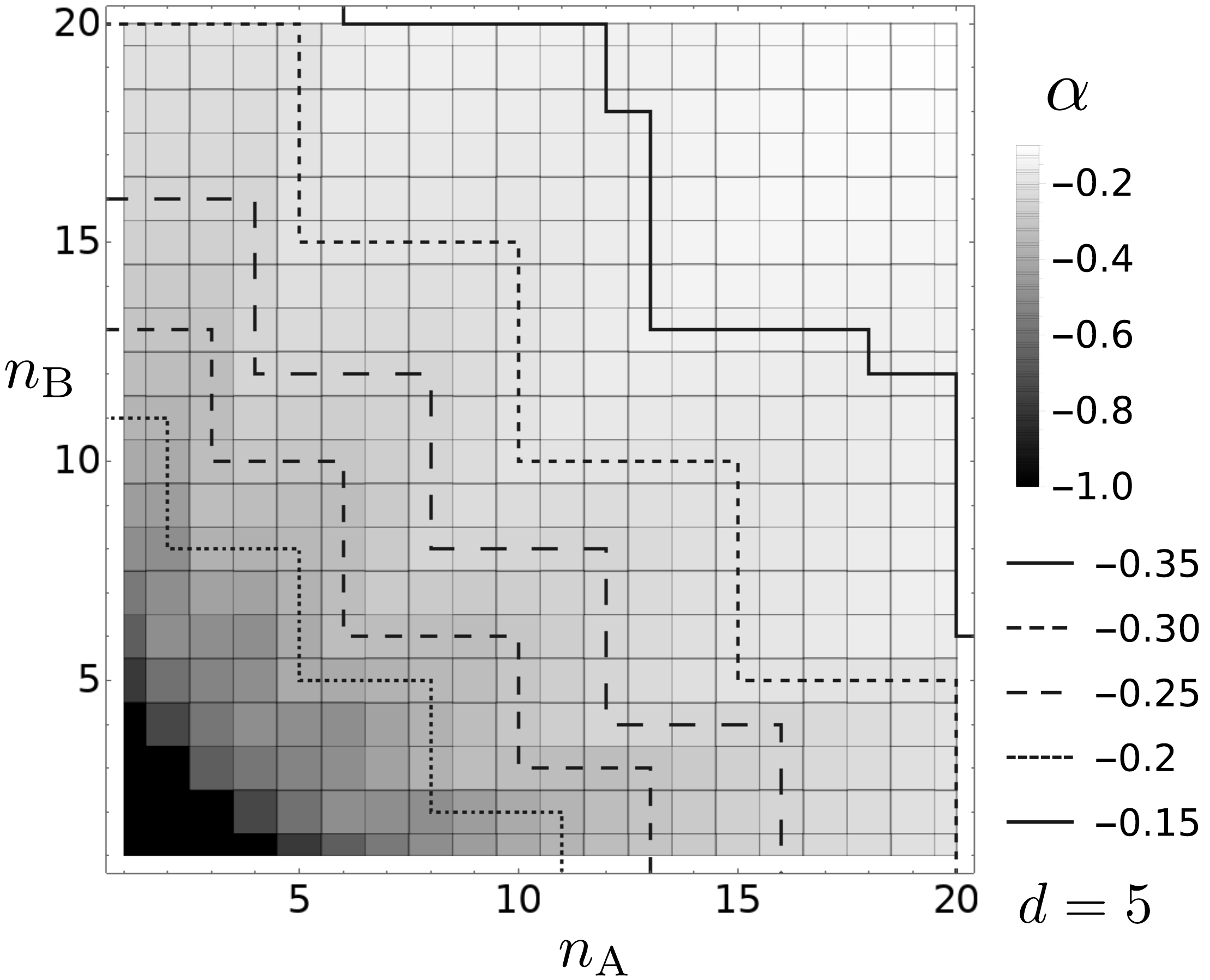}
  \caption{\label{fig:wernersol}The extreme parameters of \((\na,\nb)\)-extendible Werner states up to \(\na=\nb=20\) in the case of \(d=5\). The color temperatures in the squares represent the value of \(\alpha_{\na,\nb}\). In particular, squares with black color indicate that all Werner states are extendible for the corresponding \(\na\) and \(\nb\). These values are a consequence of the fact that the Werner state \(\rho^{\mathrm{W}}(-1)\) is a partial trace of the completely antisymmetric pure state on \(\mathcal{H}^{\otimes d}\), thus, all Werner states must be extendible for \(\na+\nb\le d\). The various lines denote the borders of the sets of \((\na,\nb)\)-pairs for which the Werner state with the corresponding \(\alpha\) parameter is extendible.}
\end{figure}

In order to obtain the extendibility, it is enough to compare the values of \(E^{\mathrm{W}}\) for the \(d-1\) pairs of diagrams in Eq.~\eqref{eq:10}. Substituting into Eq.~\eqref{eq:8} we obtain,
\begin{equation}
  \label{eq:11}
  \begin{split}
    &E^{\mathrm{W}}(\hat{\lambda}^{(\A)}(\hat{d}),\hat{\lambda}^{(\B)}(d-\hat{d}))=\\
    &\begin{cases}  -\min\left\{\frac{\hat{d}}{\na},\frac{d-\hat{d}}{\nb}\right\}\hspace{1.25cm} \text{if} \quad \left\lfloor \frac{n_\A}{\hat{d}} \right\rfloor\neq \left\lfloor \frac{\nb}{d-\hat{d}} \right\rfloor\\
       -\frac{1}{\na\nb}\left[\hat{d}(d-\hat{d})\left\lfloor \frac{\na}{\hat{d}} \right\rfloor\right.-\\
       \left.(\na\bmod \hat{d})(\nb\bmod (d-\hat{d})) \right] \text{if} \left\lfloor \frac{n_\A}{\hat{d}} \right\rfloor=\left\lfloor \frac{\nb}{d-\hat{d}} \right\rfloor
    \end{cases}
  \end{split}
\end{equation}
The modulos in the second case of Eq.~\eqref{eq:11} make it difficult to tell exactly which bipartition of \(d\) minimizes \(E^{\mathrm{W}}(\la(\hat{d}),\lb(d-\hat{d}))\), but we can reduce the number of candidates a little bit further. If we temporarily disregard the second case of Eq.~\eqref{eq:11}, the expression is minimized by choosing \(\hat{d}\) in a way that \(\hat{d}/\na\) and \((d-\hat{d})/\nb\) are the closest to each other, i.e., \(\hat{d}=\lfloor d \na/(\na+\nb)\rceil\)\footnote{In the case this value is within the bounds \(1\) and \(d-1\). Otherwise we have \(\hat{d}=1\) or \(\hat{d}=d-1\).}, where by \(\lfloor.\rceil\), we denote rounding to the closest integer. When we also take the second case of Eq.~\eqref{eq:11} into account, considering the magnitude of the term containing modulos, we get that the value of \(\hat{d}\) that minimizes \(E^{\mathrm{W}}(\la(\hat{d}),\lb(d-\hat{d}))\) differs from the one just described by at most 1. In other words, the parameter corresponding to the most entangled \((\na,\nb)\)-extendible Werner states is,
\begin{equation}
  \label{eq:12}
  \begin{split}
    &\alpha_{\na,\nb}=\min_{\hat{d}\in A\cap [1,d-1]}E^{\mathrm{W}}(\hat{\lambda}^{(\A)}(\hat{d}),\hat{\lambda}^{(\B)}(d-\hat{d})),\text{ where}\hspace{3.5cm}\\
    &A=\left\{\left\lfloor\frac{\na}{\na+\nb}\right\rceil-1, \left\lfloor\frac{\na}{\na+\nb}\right\rceil, \left\lfloor\frac{\na}{\na+\nb}\right\rceil+1 \right\}
  \end{split}
\end{equation}
We visualize this result in Figure~\ref{fig:wernersol}.

\textit{The extendibility of isotropic states.}---For the ground state problem corresponding to isotropic states, the diagram \(\lambda^{(\mathrm{LR})}\) must appear in the irrep decomposition of \(\la\otimes \overline{\lambda}^{(\B)}\).
Compared to the case of Werner states, taking the dual of Bob's \(\SU(d)\) irrep  removes the competition between  energy contributions of the terms of Eq.~\eqref{eq:4}, thus turning the Hamiltonian \(H^{\mathrm{I}}\) unfrustrated. This allows us to determine \(\beta_{\na\nb}\) with very little effort.

First, consider the case when \(\na=\nb=n\). In this situation, the eigenvalue of \(-(C_{\A}+C_{\B})\) is minimized by the same single-line diagram on both sides, \(\la=\lb=(n)\). Additionally, if we choose \(M=n\) in the definition of the dual, from Eq.~\eqref{eq:7} it is clear that \(\hat{\lambda}^{(\AB)}((n),\overline{(n)})\) is equivalent to the singlet representation that corresponds to the 0 eigenvalue of \(\tilde{C}_{\AB}\). As the quadratic Casimir operator is positive, this minimizes the energy of the \(\tilde{C}_{\AB}\) term independently of any constraint. Substituting into the eigenvalue of Eq.~\eqref{eq:4}, we obtain the extreme parameter,
\begin{equation}\label{eq:16}
  \beta_{n,n}=1+\frac{d-1}{n}.
\end{equation}
This value is equal to the previously known~\cite{johnson13_compat_quant_correl,christandl2021asymptotic} result for \((1,n)\)-extendibility. Since according to the partial order of extendibilities, \((n,1)\le(n,n')\le (n,n)\) for all \(n'\le n\), we must have \(\beta_{n,n}=\beta_{n,n'}=\beta_{n,1}\), which yields the general result,
\begin{equation}
  \label{eq:20}
\beta_{\na,\nb}=1+\frac{d-1}{\max\{\na,\nb\}}.
\end{equation}
That is, unlike Werner states, the range of extendible isotropic states has no trade-off for increasing the size of the smaller of the two subsystems to which we extend. 

\textit{Summary and Outlook}---We have determined necessary and sufficient conditions for the two-sided \((\na,\nb)\)-extendibility of Werner states for arbitrary values of \(\na\), \(\nb\) and local dimension \(d\). To achieve this result we first followed the method described in~\cite{johnson13_compat_quant_correl}, and used the symmetries of the  extendibility problem to map it into the ground state problem of a certain Hamiltonian exhibiting the same symmetries. The eigenvalues of this Hamiltonian are labeled by triples of Young diagrams that must be compatible with each other w.r.t.~the Littlewood-Richardson product of diagrams. By utilizing the dominance order of Young diagrams in the Littlewood-Richarson product, we reduced the number of variables and solved the ground state problem exactly.

We have obtained the result that the parameter range of \((\na,n_\B)\)-extendible Werner states has a non-trivial trade-off between the values of \(\na\) and \(\nb\) that depends on the divisibilities. We contrasted this with the two-sided extendibility problem of  isotropic states that, as a result of the conjugate unitary symmetry, corresponds to the ground state problem of an unfrustrated Hamiltonian. In this case, the parameter range of extendible states has no trade-off for increasing the smaller of the two extension sizes.

A straightforward direction one could further develop this result is the investigation of multipartite Werner state extendibility. In this scenario, an \(n\)-partite Werner state is shared between \(n\) composite Hilbert spaces in a permutation symmetric way. In a way analogous to our construction, it is possible to map the multipartite extendibility problem into an eigenproblem of a linear operator composed of various tensor product representations of the quadratic \(\SU(d)\) Casimir operator. The knowledge of these multipartite extendibilities could serve as a way to characterize the entanglement of multipartite Werner states.
Another possible direction would be to consider families of states with different symmetries, such as \(O\otimes O\)-symmetric bipartite states~\cite{keyl2002fundamentals,vollbrecht01_entan_measur_under_symmet}. In this case, the question of two-sided extendibility can be traced back to the fusion rules of the orthogonal group.
Finally, we hope that a generalization of our techniques could be also used to obtain the squashed entanglement of Werner states.

\noindent \textbf{Acknowledgements}
\begin{acknowledgments}
\noindent 
We thank Daniel Cavalcanti, Michał Oszmaniec and Tamás Vértesi for stimulating discussions. This work has been supported by the Ministry of Innovation and Technology and the National Research, Development and Innovation Office (NKFIH) within the Quantum Information National Laboratory of Hungary and through OTKA Grants FK 135220, K 124152, and K 124351.
\end{acknowledgments}

\bibliography{mainbibfile}

\end{document}



\title{Supplemental Material for `Extendibility of Werner States'}
\author{Dávid Jakab}
\email{jakab.david@wigner.hu}
\affiliation{Wigner Research Centre for Physics,  H-1525 Budapest P.O. Box 49, Hungary}
\affiliation{University of Pécs, Ifjúság utca 6, 7624 Pécs, Hungary}
\author{Adrián Solymos}%
\affiliation{Wigner Research Centre for Physics,  H-1525 Budapest P.O. Box 49, Hungary}
\affiliation{Eötvös Loránd University, Pázmány Péter sétány 1/A-c 1117 Budapest, Hungary}
\author{Zoltán Zimborás}
\email{zimboras.zoltan@wigner.hu}
\affiliation{Wigner Research Centre for Physics, H-1525 Budapest P.O. Box 49, Hungary}
\affiliation{Eötvös Loránd University, Pázmány Péter sétány 1/A-C 1117 Budapest, Hungary}

\maketitle
\section{Expressing the Hamiltonians with Casimir operators}
In this section, we express the linear operators,
\begin{gather}
  \label{eq:7}
  H^{\mathrm{W}}=\frac{1}{\na\nb}\sum_{i\in L,j\in R}F_{ij},\ \text{and}\\
  H^{\mathrm{I}}=-\frac{1}{\na\nb}\sum_{i\in L, j\in R}F^{t_{\B}}_{ij},
\end{gather}
with tensor product representations of the quadratic Casimir operator of \(\SU(d)\).

First, we must formulate the quadratic Casimir operator itself. We use the generators,
\begin{equation}
  \label{eq:8}
  S^{\alpha\beta}=|\beta \rangle \langle\alpha|-\frac{1}{d}\delta_{\beta\alpha}\idm,\quad \alpha,\beta=1, \ldots ,d,
\end{equation}
where \({\{ |\alpha \rangle\}}_{i=1}^{d}\) is a fixed, orthonormal basis of \(\mathbb{C}^{d}\). Since \(\sum_{i=1}^d S^{\alpha\alpha}=0\), only \(d-1\) of the \(d\) diagonal generators are independent. With our choice of generators, the quadratic Casimir element is expressed as,
\begin{equation}
  \label{eq:9}
  C=\sum_{\alpha,\beta=1}^{d}S^{\alpha\beta}S^{\beta\alpha}=\frac{d^{2}-1}{d}\idm,
\end{equation}
where the last equality is due to Schur's lemma.
The \(N\)-fold tensor product of the defining representation of \(\SU(d)\) maps \(C\) into,
\begin{equation}
  \label{eq:10}
  C_{n}=\sum_{\alpha,\beta=1}^d \left( \sum_{i=1}^n S_{i}^{\alpha\beta} \right)\left( \sum_{j=1}^n S^{\beta\alpha}_{j} \right)=2 \sum_{\alpha,\beta=1}^d \sum_{\substack{i,j=1\\i<j}}^n S_{i}^{\alpha\beta}S_{j}^{\beta\alpha}+n\frac{d^{2}-1}{d}\idm,
\end{equation}
where the lower indices \(i,j\) denote which tensor components the generators act on.

The two-fold tensor product representation of \(\SU(d)\) decomposes into two irreps, \((2,0)\) and \((1,1)\). As the flip operator \(F_{ij}\) is invariant to this representation, it can be expressed as a linear combination of two linearly independent, \(\SU(d)\) invariant two-particle operators. For this role, we choose the identity and \(\sum_{\alpha,\beta=1}^d S_{i}^{\alpha\beta}S_{j}^{\beta\alpha}\) and obtain,
\begin{equation}
  \label{eq:11}
  F_{ij}=\sum_{\alpha,\beta=1}^d S_{i}^{\alpha\beta}S_{j}^{\beta\alpha}+\frac{\idm}{d}.
\end{equation}
According to Eqs.~\eqref{eq:10} and~\eqref{eq:11}, a permutation-symmetric linear combination of flip operators involving \(n\) tensor components, such as the ones appearing in Eq.~\eqref{eq:7}, is a linear combination of \(C_{n}\) and the identity. Making use of this, we re-express \(E^{\mathrm{W}}\) as,
\begin{equation}
  \label{eq:12}
  H^{\mathrm{W}}=\frac{1}{2\na\nb}\left( C_{\AB}-C_{\A}-C_{\B}\right)+\frac{\idm}{d},
\end{equation}
where \(C_{\A}, C_{\B}\) and \(C_{\AB}\) denote \(C\) in the \(\na\), \(\nb\) and \(\na+\nb\)-fold product representations acting on Alice's and Bob's subsystems and the entire system respectively.

When in a similar fashion, we express \(H^{\mathrm{I}}\) with the generators, the partial transposition of the flip operators leads to terms proportional to \(S_{i}^{\alpha\beta}S_{j}^{\alpha\beta}\) appearing in the expression. We can relate the second, transposed generator with the dual of the defining representation, which is generated by \(-{\{S^{\beta\alpha}\}}_{\alpha,\beta=1}^{d}\). Indeed, terms of the form we are looking for appear when we express the Casimir operator in a tensor product of the defining representation and its dual. Let \(\tilde{C}_{\AB}\) denote \(C\) in the \(\na+\nb\)-fold tensor product representation in which the defining representation is used on Alice's subsystem, and its dual on Bob's subsystem,
\begin{equation}
  \label{eq:12+}
    \tilde{C}_{\AB}=\left( \sum_{k\in \A}S^{\alpha\beta}_{k}-\sum_{l\in \B} S^{\beta\alpha}  \right)\left( \sum_{k\in \A} S^{\beta\alpha}_{k}-\sum_{l\in \B} S^{\alpha\beta}_{l} \right)=
    -2\sum_{k\in \A}\sum_{l\in \B}  S_{k}^{\alpha\beta}S_{l}^{\alpha\beta}+C_{\A}+C_{\B}.
\end{equation}
Using \(\tilde{C}_{\AB}\), we can express \(H^{\mathrm{I}}\) as,
\begin{equation}\label{eq:16} H^{\mathrm{I}}=\frac{1}{2\na\nb}\left(C^{\SU(d)}_{\A}+C^{(\SU(d))}_{\B}-\tilde{C}^{\SU(d)}_{\AB}\right)+\frac{\idm}{d}.
\end{equation}

\section{The minimum product diagram}
In this section, we determine the irreducible constituent, \(\hat{\lambda}^{(\AB)}(\la, \lb)\), of the product of two arbitrary, fixed \(\SU(d)\) irreps, \(\la\otimes \lb\), that is the minimum w.r.t.~the dominance order of partitions.

In order to find the partition we are looking for, we first must look at the problem from a different angle, by fixing a different pair of \(\SU(d)\) irreps in the product. For a pair of irreps labeled by \(\la\vdash_{d}\na\) and \(\lab\vdash_{d}\na+\nb\) for which \(\la_{i}\le \lab_{i}\) for
all \(i=1,2, \ldots,d\), we define \(LR(\lab,\la)\) as the set of all \(\SU(d)\) irreps labeled by \(\lb\vdash_{d}\nb\) for which \(\lab\) is an irreducible constituent of \(\la\otimes\lb\). We further define the difference partition \(\lab-\la\vdash_{d}\nb\) as the integer partition created from the pointwise difference of the two partitions by sorting its elements into decreasing order,
\begin{equation}
  \label{eq:2}
  \lab-\la=\text{sort}{\{\lab_{i}-\la_{i}\}}_{i=1}^{d}.
\end{equation}

By using the symmetry properties of the skew-diagrams that appear in the Littlewood-Richardson algorithm describing the fusion rules of \(\SU(d)\), in Ref.~\cite{azenhas99_admis_inter_invar_factor_produc_matric} it was
shown that \(\lab-\la\in LR(\lab,\la)\), furthermore, for all \(\lb\in LR(\lab,\la)\),
\begin{equation}
  \label{eq:1}
  \lab-\la\unlhd \lb.
\end{equation}

In order to obtain from this result, the partition \(\hat{\lambda}^{(\AB)}(\la, \lb)\), we make use of the dual symmetry of the of the fusion rules of \(\SU(d)\). The irrep decomposition of \(\la\otimes \lb\) is described by the multiplicities \(m(\la,\lb,\lab)\) in,
\begin{equation}
  \label{eq:3}
  \la\otimes \lb=\bigoplus_{\lab}m(\la,\lb,\lab)\lab.
\end{equation}
These multiplicities are invariant to exchanging \(\lab\) and \(\lb\), then replacing both representations with their respective duals. This symmetry is evident if one expresses the multiplicities using the inner product of characters,
\begin{equation}
  \label{eq:4}
  \begin{split}
    m(\la,\lb,\lab) = \langle \chi_{\la}\chi_{\lb}, \chi_{\lab} \rangle= \int_{\SU(d)}  \chi_{\la}(u)\chi_{\lb}(u)\overline{\chi}_{\lab}(u)\mathrm{d}u\\
    =\int_{\SU(d)} \chi_{\la}(u)\chi_{\overline{\lambda}^{(\AB)}}(u)\overline{\chi}_{\overline{\lambda}^{(\B)}}{(u)}\mathrm{d}u=m(\la,\overline{\lambda}^{(\AB)}\overline{\lambda}^{(\B)}),
  \end{split}
\end{equation}
where by \(\chi_{\lambda}\), we denote the character of an irrep \(\lambda\), and the integration is over the invariant Haar measure of \(\SU(d)\).

The exchange of \(\lab\) and \(\lb\) in Eq.~\eqref{eq:4} gives us a way to obtain the minimum product diagram from Eq.~\eqref{eq:1}, but first, we must make sure that the operation of mapping an irrep to its dual does not influence dominance order. If \(\lambda,\mu\vdash_{d} n\), \(\lambda\unrhd \mu\) and we take the \(M\)-dual of both partitions, so as to make sure that dominance order is well defined between the results, then for all \(1\le l\le d\),
\begin{equation}
  \label{eq:7+}
    n-(d-l)M+\sum_{i=1}^{d-l}\overline{\mu}=
    n-\sum_{i=l+1}^d\mu_{i} =\sum_{i=1}^l \mu_{i}\le \sum_{i=1}^l \lambda_{i}=n-\sum_{i=l+1}^d \lambda_{i}=n-(d-l)M+\sum_{i=1}^{d-l}\overline{\lambda};
\end{equation}
therefore, \(\overline{\lambda}\unrhd \overline{\mu}\).

From combining Eqs.~\eqref{eq:1} and~\eqref{eq:4} we obtain,
\begin{equation}
  \label{eq:5}
  \lab-\la =\min\{\lb\vdash_{d}\nb : m(\la,\lb,\lab)>0\}=\min\{\lb\vdash_{d}\nb : m({\la,\overline{\lab}, \overline{\lb}})>0\}.
\end{equation}
As the dual does not influence dominance order, relabeling the partitions gives us the result we were looking for,
\begin{equation}
  \label{eq:6}
  \hat{\lambda}^{(\AB)}(\la, \lb)=\min\{\lab\vdash_{d}\na+n'_{\B}:m(\la,\lb,\lab)>0\}=\overline{(\overline{\lambda}^{(\B)}-\la)}=\text{sort}{\{\la_i+\lb_{d-i+1}\}}_{i=1}^{d},
\end{equation}
where \(n'_{\B}=Md-\nb\) and \(\lambda^{\B}\vdash_{d}n'_{\B}\). In other words, to obtain the minimum product diagram, one has to simply attach the diagram \(\lb\) upside-down to the left-side of \(\la\), and sort the resulting shape by its row length into a proper Young diagram.

\section{The ground state problem of Werner states}
In this section, we show that the pair of partitions that minimizes the eigenvalue of \(H^{\mathrm{W}}\) from Eq.~\eqref{eq:12} that corresponds to the triple of \(\SU(d)\) irreps \((\la,\lb, \hat{\lambda}^{(\AB)}(\la, \lb))\); that is, the \(\la,\lb\) pair that minimizes the expression,
\begin{equation}
  \label{eq:20}
  E^{\mathrm{W}}(\la,\lb)= \frac{1}{\na\nb}\sum_{i=1}^d \left[ \la_{i}\lb_{d-i+1}-i(\text{sort}{({\{\la_{j}+\lb_{d-j+1}\}}_{j=1}^{d})}_{i}\right.- \left.(\la_{i}+\lb_{i})) \right],
\end{equation}
is composed of the dominance order minima of all integer \(\hat{d}\)-partitions of \(\na\) and \(d-\hat{d}\)-partitions of \(\nb\) for some \(\hat{d}=1,2, \ldots d-1\). In other words,
\begin{equation}
  \label{eq:24}
  \min_{\la\vdash_{d}\na, \lb\vdash_{d}\nb}E^{\mathrm{W}}(\la,\lb)=E^{\mathrm{W}}(\hat{\lambda}^{(\A)}(\hat{d}),\hat{\lambda}^{(\B)}(d-\hat{d})),
\end{equation}
where,
\begin{equation}
  \label{eq:25}
  \begin{split}
    &\hat{\lambda}^{(\A)}(d_\A)=\left(\left\lceil\frac{\na}{d_\A}\right\rceil, \left\lceil\frac{\na}{d_\A}\right\rceil, \ldots , \overset{\na\bmod d_\A\text{-th}}{\left\lceil\frac{\na}{d_\A}\right\rceil}\right.,
      \left.\left\lfloor \frac{\na}{d_\A} \right\rfloor,\left\lfloor \frac{\na}{d_\A} \right\rfloor, \ldots , \overset{d_\A\text{-th}}{\left\lfloor \frac{\na}{d_\A} \right\rfloor}\right),\\
    &\hat{\lambda}^{(\B)}(d_{\B})=\left(\left\lceil\frac{\nb}{d_\B}\right\rceil,\left\lceil\frac{\nb}{d_\B}\right\rceil, \ldots , \overset{\nb\bmod d_\B\text{-th}}{\left\lceil\frac{\nb}{d_\B}\right\rceil}\right., \left.\left\lfloor \frac{\nb}{d_\B} \right\rfloor,\left\lfloor \frac{\nb}{d_\B} \right\rfloor, \ldots,\overset{d_\B\text{-th}}{\left\lfloor \frac{\nb}{d_{\B}} \right\rfloor}\right).
  \end{split}
\end{equation}
To prove our statement, we will create a path of diagram pairs starting from an arbitrarily chosen \(\la\vdash_{d} \na\) and \(\lb\vdash_{d}\nb\), that terminates in one of the pairs in Eq~\eqref{eq:25}, along which the value of \(E^{\mathrm{W}}\) is weakly decreasing.

\begin{figure}[ht!]
  \centering
  \includegraphics[width=0.45\textwidth]{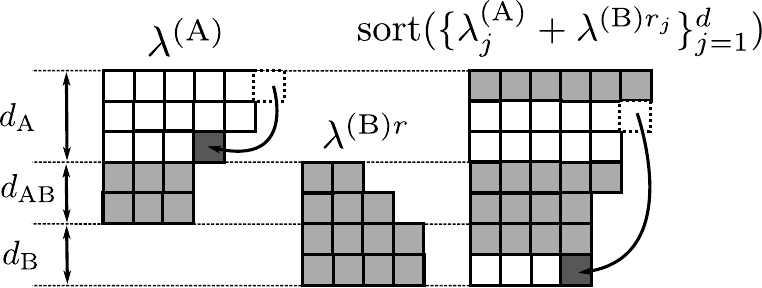}
  \caption{\label{fig:wernersolution1} An example for a transformation that decreases the non-overlapping part of \(\la\) in dominance order. The distance between the two affected rows in \(\text{sort}({\{\la_{j}+\lambda^{(\B)r}_{j}\}}_{j=1}^{d})\) is always larger than or equal to that in \(\la\), as the sorting can potentially shuffle in additional rows between the ones present in \(\la\). These are indicated with a darker color.}
\end{figure}

We denote the reversal of a partition \(\lambda\) with \(\lambda^{r}\), i.e., \(\lambda^{r}_{i}=\lambda_{d-i+1}\); furthermore, we denote the number of overlapping rows between \(\la\) and \(\lambda^{(\B)r}\) by \(d_{\AB}\), and the numbers of non-overlapping rows of \(\la\) and \(\lambda^{(\B)r}\) by \(d_{\A}\) and \(d_{\B}\) respectively. Therefore, the number of rows of \(\la\) is \(d_{\A}+d_{\AB}\le d\) and that of \(\lb\) is \(d_{\B}+d_{\AB}\le d\), where \(d_{\A}+d_{\B}+d_{\AB}=d\) and \(\la_{i}\lambda^{(\B)r}_{i}\neq 0\) iff \(d_{\A}+1\le i\le d_{\A}+d_{\AB}\), see Figure~\ref{fig:wernersolution1}. Moreover, we denote the number of boxes in the non-overlapping parts by  \(n'_{\A}=\sum_{i=1}^{d_{\A}}\la_{i}\) and \(n'_{\B}=\sum_{i=1}^{d_{\B}}\lb_{i}\). We build the path from two different types of steps. In the first one, we transform the non-overlapping parts of the two diagrams into a standard form. In the second one, we take a pair of diagrams for which the non-overlapping parts are in the standard form and move boxes from the overlapping into the non-overlapping parts.

Consider the transformation that consists of moving a single box downward within the non-overlapping part of \(\la\) in a way that results in a valid integer partition. That is, we transform \(\la\) into \(\lambda^{(\A)'}\), where \(\lambda^{(\A)'}_{i}=\la_{i}-1\), \(\lambda^{(\A)'}_{j}=\la_{j}+1\) for some \(1\le i<j\le d_{\A}\) and all other rows of \(\la\) stay unchanged, see Figure~\ref{fig:wernersolution1}. An important detail to take note of here, is that we can always choose the order of rows of \(\text{sort}({\{\la_{j}+\lambda^{(\B)r}_{j}\}}_{j=1}^{d})\) in a way to make it invariant to the transformation. If there is ambiguity in the order, i.e., \({\{\la_{j}+\lambda^{(\B)r}_{j}\}}_{j=1}^{d}\) has multiple elements equal to \(\la_{i}\) or \(\la_{j}\), we choose \(\la_{i}\) to be the bottommost, and \(\la_{j}\) to be the topmost of them.

Let us compute the change in \(E^{\mathrm{W}}\) after transforming \(\la\) into \(\lambda^{(\A)'}\).  Since the overlapping part stays unchanged, the quadratic term of Eq.~\eqref{eq:20} has no contribution. The contribution of the remaining terms depends on the distance between the two changed rows in \(\la\), and in \(\text{sort}({\{\la_{j}+\lambda^{(\B)r}_{j}\}}_{j=1}^{d})\),
\begin{multline}\label{eq:26}
  E^{\mathrm{W}}(\lambda^{(\A)'},\lb)-E^{\mathrm{W}}(\la,\lb)=\\
  \frac{1}{\na\nb}\left[j-i-\right. \left.\left(\text{first}(\la_{j},\text{sort}({\{\la_{j}+\lambda^{(\B)r}_{j}\}}_{j=1}^{d})) - \text{last}(\la_{i},\text{sort}({\{\la_{j}+\lambda^{(\B)r}_{j}\}}_{j=1}^{d}))\right)\right]\le0,
\end{multline}
where \(\text{first}(x,y)\) and \(\text{last}(x,y)\) denote the first and last positions of \(x\) in the sequence \(y\). The distance between the last occurrence of \(\la_{i}\) and the first occurrence of \(\la_{j}\) in \(\text{sort}({\{\la_{j}+\lambda^{(\B)r}_{j}\}}_{j=1}^{d})\) must be at least \(j-i\), since \(\la_{i+1},\la_{i+2}, \ldots ,\la_{j-1}\) all lie between the two elements. Additionally, since \(\text{sort}({\{\la_{j}+\lambda^{(\B)r}_{j}\}}_{j=1}^{d})=\text{sort}({\{\lambda^{(\A)r}_{j}+\lb_{j}\}}_{j=1}^{d})\), performing the analog of this transformation on \(\lb\) weakly decreases \(E^{\mathrm{W}}\) as well.

Repeating the transformation just described on both \(\la\) and \(\lb\), creates sequences of diagrams that are strictly decreasing in dominance order. As the sets \(\{\lambda\vdash_{d_{\A}}n'_{\A}\}\) and \(\{\lambda\vdash_{d_{\B}}n'_{\B}\}\) both have a minimum element, continuing until there is no legal way left to move boxes downwards within the non-overlapping parts of \(\la\) and \(\lb\), eventually transforms these parts into the these minimum elements. That is, the end results of the repeated transformations are,
\begin{equation}\label{eq:27}
  \begin{split}
    &\hat{\lambda}^{(\A)''}=\left(\left\lceil\frac{n'_{\A}}{d_{\A}}\right\rceil, \left\lceil\frac{n'_{\A}}{d_{\A}}\right\rceil, \ldots , \overset{n'_{\A}\bmod d_{\A}\text{-th}}{\left\lceil\frac{n'_{\A}}{d_{\A}}\right\rceil}\right.,
      \left.\left\lfloor \frac{n'_{\A}}{d_{\A}} \right\rfloor,\left\lfloor \frac{n'_{\A}}{d_\A} \right\rfloor, \ldots , \overset{d_\A\text{-th}}{\left\lfloor \frac{n'_{\A}}{d_\A} \right\rfloor}, \la_{d_{\A}+1}, \la_{d_{\A}+2}, \ldots \la_{d}\right),\\
    &\hat{\lambda}^{(\B)''}=\left(\left\lceil\frac{n'_{\B}}{d_{\B}}\right\rceil,\left\lceil\frac{n'_{\B}}{d_{\B}}\right\rceil, \ldots , \overset{n'_{\B}\bmod d_{\B}\text{-th}}{\left\lceil\frac{n'_{\B}}{d_{\B}}\right\rceil}\right., \left.\left\lfloor \frac{n'_{\B}}{d_{\B}} \right\rfloor,\left\lfloor \frac{n'_{\B}}{d_{\B}} \right\rfloor, \ldots,\overset{d_{\B}\text{-th}}{\left\lfloor \frac{n'_{\B}}{d_{\B}} \right\rfloor}, \lb_{d_{\B}+1}, \lb_{d_{\B}+2}, \ldots , \lb_{d}\right).
  \end{split}
\end{equation}
In this way, we are able to transform any pair of diagrams into this standard form without increasing \(E^{\mathrm{W}}\).

\begin{figure}[ht]
  \centering
  \includegraphics[width=0.56\textwidth]{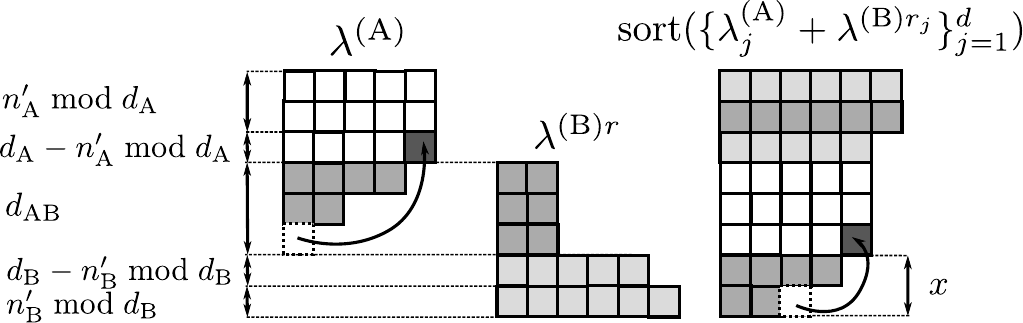}
  \caption{\label{fig:wernersolution2} A transformation of a pair of diagrams of the standard form described in Eq.~\eqref{eq:27}, that moves a box form the bottom row of \(\la\) into the non-overlapping part. When \(\la\) and \(\lb\) both have the standard form, the diagram \(\text{sort}({\{\la_{j}+\lambda^{(\B)r}_{j}\}}_{j=1}^{d})\) is composed of four rectangular sections of width \(\na'\bmod d_{\A}\), \(\na'\bmod d_{\A}-1\), \(\nb'\bmod d_{\B}\) and \(\nb'\bmod d_{\B}-1\) respectively. Additionally, the \(d_{\AB}\) overlapping rows are shuffled between these rectangular sections in an unknown way.
    In the diagram \(\la\), \(d_{\AB}-1\), overlapping rows are between the original and new positions of the box, while in \(\text{sort}({\{\la_{j}+\lambda^{(\B)r}_{j}\}}_{j=1}^{d})\), these overlapping rows are not necessarily between the two positions. }
\end{figure}

We define a second type of transformation that acts on pairs of diagrams, \(\la, \lb\) of the form described in Eq.~\eqref{eq:27}; we further assume that \(d_{\AB}>0\). Consider the transformation that takes a single box from the bottommost overlapping row of \(\la\), and attaches it to the non-overlapping part in the way that makes the resulting diagram the smallest w.r.t.\ dominance order. I.e., we transform \(\la\) to \(\lambda^{(\A)'}\) where \(\lambda^{(\A)'}_{d_{\A}+d_{\AB}}=\la_{d_{\A}+d_{\AB}}-1\), \(\lambda^{(\A)'}_{n'_{\A}\bmod d_{\A}+1}=\la_{n'_{\A}\bmod d_{\A}+1}+1\), and all other rows stay unchanged, see Figure~\ref{fig:wernersolution2}.

When our transformation moves the box downwards in \(\text{sort}({\{\la_{j}+\lambda^{(\B)r}_{j}\}}_{j=1}^{d})\), the contribution of every term in \(E^{\mathrm{W}}(\lambda^{(\A)'},\lb)-E^{\mathrm{W}}(\la,\lb)\) is non-positive; therefore, we need only be concerned with the cases in which the box moves upwards. Without loss of generality, we can assume that \(n'_{\A}/d_{\A}\le n'_{\B}/d_{\B}\), thus, we move the box into the top row of the bottommost rectangular section of \(\text{sort}({\{\la_{j}+\lambda^{(\B)r}_{j}\}}_{j=1}^{d})\). In the case of the contrary, we simply apply the transformation to \(\lb\) instead of \(\la\). This means, that the only arrangements in which the box moves upwards, are the ones where it is taken from one of those rows that are shuffled below all the non-overlapping rows. Let us assume that we take the box from the \(x\)-th row below the bottommost non-overlapping row, i.e., it moves \(x+d_{\A}-n'_{\A}\bmod d_{\A}-1\) rows upward in \(\text{sort}({\{\la_{j}+\lambda^{(\B)r}_{j}\}}_{j=1}^{d})\), and \(d_{\AB}+d_{\A}-n'_{\A}\bmod d_{\A}-1\) upwards in \(\la\). This way, the change in \(E^{\mathrm{W}}\) is,
\begin{equation}\label{eq:28}
  E^{\mathrm{W}}(\lambda^{(\A)'},\lb)-E^{\mathrm{W}}(\la,\lb)=-\lb_{d_{\B}+1}+x-d_{\AB}\le0.
\end{equation}

Starting from any pair of diagrams, by first rearranging the non-overlapping rows into the standard form in Eq.~\eqref{eq:27}, then moving the boxes from the overlapping rows into the non-overlapping ones with the method just described, we eventually reach one of the diagrams in Eq.~\eqref{eq:25} without increasing the value of \(E^{\mathrm{W}}\).

\bibliography{mainbibfile}